\documentclass[12pt]{article}
\usepackage[english]{babel}
\usepackage[utf8]{inputenc}
\usepackage[T1]{fontenc}

\usepackage{amsmath}
\def\barT{\bar{T}}

\usepackage{amsfonts}

\usepackage{url}
\usepackage[dvips]{graphicx}
\usepackage{calc}
\usepackage{subfigure}
\usepackage{multirow}

\def\tX{\tilde{X}}

\begin{document}
	\begin{titlepage}
	\begin{center}
		{\Large{ \bf Remark About Action for Rastall Gravity}}
		
		\vspace{1em}  
		
		\vspace{1em} J. Kluso\v{n} 			
		\footnote{Email address:
			klu@physics.muni.cz (J.
			Kluso\v{n}) }\\
		\vspace{1em}
		\textit{Department of Theoretical Physics and
			Astrophysics, Faculty of Science,\\
			Masaryk University, Kotl\'a\v{r}sk\'a 2, 611 37, Brno, Czech Republic}
		
		\vskip 0.8cm
		
		%
		%
		%
		%
		%
		%
		
		\vskip 0.8cm
		
	\end{center}

\begin{abstract}
In this short note we study Rastall gravity action as was proposed by Golovnev in \cite{Golovnev:2023evp}. We construct its reparametrization invariant form and determine corresponding equations of motion.
\end{abstract}

\bigskip

\end{titlepage}

\newpage

\section{Introduction and Summary}\label{first}
Modified theories of gravity have been studied for many years for several reasons. One important reason is their  potential phenomenological applications since they provide hints on how to solve
current cosmological problems, as Hubble tension or mistery of cosmological constant. Another argument for studying 
modified theories is purely theoretical, since we would like to see whether it is possible to construct generalized theories of gravity  whose structure is beyond the standard Enstein-Hilbert action 
\cite{Shankaranarayanan:2022wbx,Nojiri:2017ncd}. A famous example of modified theories of gravity  is $F(R)$ gravity where an action 
is general function of scalar curvature  $F(R)$
\cite{DeFelice:2010aj}.  Another example of an interesting modification of gravity 
are Dirac-Born-Infeld inspired theories of gravity \cite{BeltranJimenez:2017doy}.

Lesser-known  example of an alternative theory of gravity is the so-called
Rastall gravity \cite{Rastall:1972swe}, where the basic premise is the denial  of the presumption of conservation of matter stress energy tensor. Remarkably, this idea became popular in recent years with number of papers that discuss its phenomonological consequences, see for example 
\cite{Vitohekpon:2026nlb,Nogales:2026cpr,Herrera:2025duu,Errehymy:2024lhl,Afshar:2023uyw,Tangphati:2023nwz,Shahidi:2021lxt,Darabi:2017coc}. This fact is even more remarkable when we think deeply about theoretical arguments for an existence of this theory. In more
details, it was nicely discussed in 
\cite{Golovnev:2023evp} that it is very difficult to implement non-conservation of stress energy tensor since matter part of the action is diffeomorphism invariant and hence thanks to the Noether theorem stress energy tensor is conserved when the matter equations of motion  are satisfied. On the other hand such a non-conservation of the stress energy 
tensor can be implemented at the level of gravity equations of motion when the relative coefficients between Ricci tensor and scalar curvature are changed. 
 However then it was  shown in 
\cite{Visser:2017gpz} that Rastall gravity equations of motion can be rewritten into the form of General relativity equations of motion by simple rearrangement of the matter content of theory. In other words there is nothing new on the gravity side in case of Rastall gravity.

Similar discussion was then extended and further clarified in 
\cite{Golovnev:2023evp} where Rastall gravity action was also
proposed. A characteristic property of this action is that
it has the form of linear combination of general and unimodular gravity \cite{Unruh:1988in,Kuchar:1991xd,Henneaux:1989zc}.
It is well known that unimodular gravity action 
is not invariant under full diffeomorphism due to the presence
of term $\sqrt{-g}-1$. Then it was  shown in \cite{Kuchar:1991xd}   that  manifest covariance of unimodular theory can be restored by reparametrization procedure. Since Rastall
gravity action as was proposed in \cite{Golovnev:2023evp} breaks diffeomorphism invariance as well, it is natural to find 
its manifest diffeomorphism invariant form. Exactly this is the goal of this short note and we obtain manifestly diffeomorphism invariant action for Rastall gravity using deparametrization 
procedure \cite{Kuchar:1991xd}. We mean that such an action  is natural   starting point for further analysis since all physical theories should   be formulated in such a way that they do not depend on preferred set of coordinates. It is important to stress that this diffeomorphism invariant action now contains divergence of four vector field $\tau^\mu$
whose components have to be considered as dynamical degrees of freedom. Consequently    their equations of motion should be determined from covariant Rastall
action by variation with respect to $\tau^\mu$.

 If we derive these equations we find new condition 
which is absent when the Rastall action was written in unimodular
coordinates. Explicitly, this condition states that scalar curvature has to be constant and this condition holds for arbitrary form of matter. 
This fact restricts possible solutions of Rastall gravity in a significant way. In other words equations of motion derived
from covariant form of Rastall gravity action are not fully 
equivalent to postulated Rastall gravity equations of motion.
Then we can say that 
 Rastall gravity action  as was proposed in \cite{Golovnev:2023evp} has certain limitation and proper action for Rastall gravity has not been found yet. However question 
 is whether we should search for such an action since
\cite{Visser:2017gpz,Golovnev:2023evp} clearly showed that Rastall 
gravity is equivalent to General Relativity.

\section{Brief Review of Rastall Gravity}
Rastall gravity is defined by simple change of coefficients
in front of $R_{\mu\nu}$ and $R$ in Einstein equations of motion. 
Explicitly we can write the Rastall equations of motion 
as 
\begin{equation}\label{Rastalleq}
	R_{\mu\nu}+\left(l-\frac{1}{2}\right)Rg_{\mu\nu}=T^R_{\mu\nu} \ , 
\end{equation}
where $T_{\mu\nu}^R$ we call Rastall energy momentum tensor.
Now we briefly review arguments presented in
 \cite{Visser:2017gpz,Golovnev:2023evp} that Rastall gravity 
 is equivalent to General Relativity.

First of all we multiply (\ref{Rastalleq}) with $g^{\mu\nu}$
and we obtain
\begin{equation}
	R=\frac{T^R}{4l-1} \ , \quad T^R=g^{\mu\nu}T_{\mu\nu}^R \ . 
\end{equation}
We see at  that value $l=\frac{1}{4}$ is special. For the time being we will presume that $l\neq \frac{1}{4}$
so that we can rewrite the action (\ref{Rastalleq}) as
\begin{equation}
R_{\mu\nu}-\frac{1}{2}Rg_{\mu\nu}\equiv G_{\mu\nu}=
T_{\mu\nu}^R-\frac{l}{4l-1}T^Rg_{\mu\nu}\equiv \barT_{\mu\nu} \ . 
\end{equation}
Due to the fact that Einstein tensor obeys identity $\nabla^\mu G_{\mu\nu}$ we immediately see from the equation above that 
the stress energy tensor $\barT_{\mu\nu}$ is conserved
\cite{Visser:2017gpz,Golovnev:2023evp}
\begin{equation}
	\nabla^\mu \barT_{\mu\nu}=0 \ . 
\end{equation}
In other words  the dynamics of gravity is the same as in General relativity where the only modification appears in the definition of $\barT_{\mu\nu}$. It is important
to stress that $\barT_{\mu\nu}$ is conserved and hence should be considered
as physical one. 
Another important question related to Rastall gravity is that its formulation was
performed on the level of equations of motion since it was thought for long time 
that  an action for Rastall gravity is not known. However such an action was proposed recently in \cite{Golovnev:2023evp}. In the next section we
will review its properties. 
\section{ Rastall Gravity Action}
In  \cite{Golovnev:2023evp} the action for Rastall gravity was proposed in the form 
\begin{equation}\label{Rastallact}
S=\frac{1}{1+c}\int d^4X
(\sqrt{-g}[R+\zeta (\beta-R)]+c\beta+\lambda(
\sqrt{-g}-1)^2) \ , 
\end{equation}
where $\zeta,\beta$ and $\lambda$ are auxiliary fields that 
are scalars by definition and where $c$ is an arbitrary constant.  Further, $R$ is scalar curvature of the metric $g_{\mu\nu}$. Finally $\lambda$ is Lagrange multiplier that ensures that the metric obeys unimodular condition $\sqrt{-g}=1$ on shell. Note that due to the presence of terms $\beta R$ and $\lambda$ the action 
(\ref{Rastallact}) is not invariant under full diffeomorphism transformations. This is crucial fact as we will see below. 

Let us now show how the equations of motion derived  from 
(\ref{Rastallact}) lead to Rastall gravity equations of motion.
 Using well known formula for variation of scalar curvature in the form 
\begin{equation}
\delta R=R_{\mu\nu}\delta g^{\mu\nu}-\nabla_\mu \nabla_\nu \delta g^{\mu\nu}+g_{\mu\nu}\Box \delta g^{\mu\nu} \  
\end{equation}
we obtain that the variation of the action (\ref{Rastallact}) 
with respect to  $\delta g^{\mu\nu}$ gives
\begin{eqnarray}\label{eqg}
(1-\zeta)G_{\mu\nu}+ (\nabla_\mu\nabla_\nu \zeta-g_{\mu\nu}\Box \zeta )
	-\frac{1}{2}
g_{\mu\nu}\zeta \beta-\lambda g_{\mu\nu}
	(\sqrt{-g}-1)=0 \ ,  \nonumber \\
\end{eqnarray}
where $G_{\mu\nu}=R_{\mu\nu}-\frac{1}{2}Rg_{\mu\nu}$
and where
$\nabla_\mu$ is the covariant derivative compatible with $g_{\mu\nu}$ and $\Box=\nabla_\rho\nabla^\rho$.

Further, equations of motion for $\zeta,\beta$ and $\lambda$ have the form 
\begin{eqnarray}\label{eqmaux}
\beta-R=0 \ , \quad \zeta \sqrt{-g}+c=0 \ , \quad 
\sqrt{-g}-1=0 \ . 
\nonumber \\
\end{eqnarray}
If we combine the second and the third equation above we immediately get that $\zeta=-c$. Inserting this result together with $\beta=R$
 into 
(\ref{eqg}) we obtain 
\begin{equation}
(1+c)G_{\mu\nu}+\frac{1}{2}g_{\mu\nu}cR=0 
\end{equation}
or equivalently
\cite{Golovnev:2023evp}
\begin{equation}\label{Rastall}
	R_{\mu\nu}+(l-\frac{1}{2})g_{\mu\nu}R=0 \ ,  \quad 
	l=\frac{c}{2(1+c)} \ 
\end{equation}
that has the form of equations of motion of Rastall gravity with absence of matter. 
This result nicely demonstrates that  equations of motion of Rastall gravity can be derived by variation principle
from the action  (\ref{Rastallact}).

Before we proceed further it is instructive to show why it is necessary that the action depends on the combination $\sqrt{-g}-1$ quadratically
\cite{Golovnev:2023evp}. Let us for the time
being presume linear dependence on 
 $\sqrt{-g}-1$ so that  the equations of motion have the form 
\begin{eqnarray}
&&	(1-\zeta)G_{\mu\nu}+ (\nabla_\mu\nabla_\nu \zeta-g_{\mu\nu}\Box \zeta )
	-\frac{1}{2}
	g_{\mu\nu}\zeta \beta-\frac{1}{2}\lambda g_{\mu\nu}
=0 \ ,\nonumber \\
&&	\beta-R=0 \ , \quad \zeta \sqrt{-g}+c=0 \ , \quad 
	\sqrt{-g}-1=0 \ . 
	\nonumber \\
\end{eqnarray}
We again have that $\zeta=-c$ so that the equations of motion for $g_{\mu\nu}$ reduce into
\begin{equation}\label{equni}
	(1+c)G_{\mu\nu}+\frac{c}{2}g_{\mu\nu}R-\frac{1}{2}\lambda g_{\mu\nu}=0 \ . 
\end{equation}
Taking the trace of the equation above  we get that $\lambda$ is equal to
\begin{eqnarray}
\lambda=\frac{1}{2}(c-1)R \ . 
\end{eqnarray}
Inserting this result back to (\ref{equni})  we obtain 
\begin{equation}
(1+c)(R_{\mu\nu}-\frac{1}{4}Rg_{\mu\nu})=0
\end{equation}
which is equation of motion of unimodular gravity. Note that  the same equation 
can be derived from (\ref{Rastall}) when $l=\frac{1}{4}$.
\section{Deparametrization of Rastall Gravity Action}\label{second}
We demonstrated in previous section that the action (\ref{Rastallact}) really leads to the equation of motion of Rastall gravity. On the other hand this action manifestly breaks
diffeomorphism invariance due to the presence of terms 
$c \beta $ and $(\sqrt{-g}-1)^2$. It is basic postulate of theoretical physics that action 
functional  defined on four dimensional space-time should be invariant under reparametrizations. Say differently, action functional has to be formulated in such a way that there is no special set of coordinates. It is remarkable that here is a nice procedure how makes any action invariant under diffeomorphism transformation
which is known as deparametrization \cite{Kuchar:1991xd} and we apply this procedure to the action  (\ref{Rastallact}).
Let us denote coordinates that are used in (\ref{Rastallact})
 as  $X^A=(T,Z^k)$ and we call them as generalized unimodular coordinates. Then following \cite{Kuchar:1991xd}
 we can restore full diffeomorphism
invariance by parametrization. The privileged generalized unimodular variables $X^A$ are replaced
with arbitrary coordinates $x^\mu$ and are promoted to the field variables $X^A=X^A(x^\mu)$.  As the next step  we introduce Jacobian of transformation from $x^\mu\rightarrow X^A$
\cite{Kuchar:1991xd}
\begin{equation}
	\tX=\frac{1}{4!}
	\epsilon_{ABCD}X^A_\alpha X^B_\beta X^C_\gamma X^D_\delta \epsilon^{\alpha\beta\gamma\delta} \quad X^A_\alpha=\frac{\partial X^A}{\partial x^\alpha} \ .
\end{equation}
Note that under diffeomorphism transformations $x'^\mu=f^\mu(x)$ we have that $X^A$ transform as scalars 
\begin{equation}
	X'^A(x')=X^A(x) \ 
\end{equation}
so that we immediately get that $\tX$ transforms as 
\begin{equation}	
	\tX'(x')=
	\tX(x) \frac{1}{J} \ , \quad J\equiv \det J_\beta^{\alpha'} \ , \nonumber \\
\end{equation}
where we introduced $4\times 4 $ matrix $J^{\alpha'}_{\beta}$ 
\begin{equation}
	J^{\alpha'}_{\beta}=\frac{\partial x'^{\alpha'}}{\partial x^\beta}
	\ , \quad 
	(J^{-1})^{\alpha}_{\beta'}=
	\frac{\partial x^\alpha}{\partial x^{\beta'}}
\end{equation}
that obey relations
\begin{equation}
	J^{\alpha'}_{\beta}(J^{-1})^\beta_{\gamma'}=
	\delta^{\alpha'}_{\gamma'} \ , 
	\quad (J^{-1})^{\alpha}_{\beta'}J^{\beta'}_{\beta}=
	\delta^\alpha_\beta \ .
\end{equation}
Note that than we have
\begin{eqnarray}
&&	\int d^4x'\beta'\tX'=
\int d^4x \beta \tX \ , \nonumber \\
	\nonumber \\
&&\int d^4x'\lambda'(	\tX'-\sqrt{-g'})=
\int d^4x \lambda(\tX-\sqrt{-g}) \nonumber \\
\end{eqnarray}
using the fact that $d^4x'=J d^4x$.
Finally note that we can write $\tX$ as \cite{Kuchar:1991xd}
\begin{eqnarray}
	\tX=
	\partial_\alpha \tau^\alpha \ , \quad 
	\tau^\alpha=\frac{1}{3!}\epsilon_{IJK}T Z^I_\beta Z^J_\gamma
	Z^K_\delta \epsilon^{\alpha\beta\gamma\delta} \ , 
	\nonumber \\
\end{eqnarray}
where $X^A=(T,Z^K), I,J,K=1,2,3$. 

Using these results we can formulate  procedure how to make the action (\ref{Rastallact}) diffeomorphism invariant. We introduce factor $\partial_\alpha \tau^\alpha$ in front of the scalar $\beta$ and replace $\sqrt{-g}-1$ with 
$\sqrt{-g}-\partial_\alpha\tau^\alpha$. Further, since the action depends on this combination quadratically we have to multiply it with prefactor $\frac{1}{\partial_\alpha\tau^\alpha}$. Collecting
these terms together we obtain diffeomorphism invariant action 
for Rastall gravity in the form 
\begin{equation}\label{Sfinal}
	S_R=\frac{1}{1+c}\int d^4x
	(\sqrt{-g}[R+\zeta (\beta-R)]+c\beta \partial_\alpha \tau^\alpha+\frac{1}{\partial_\alpha \tau^\alpha}\lambda(
	\sqrt{-g}-\partial_\alpha \tau^\alpha)^2) \ . 
\end{equation}
This diffeomorphism invariant form of Rastall action is the key result of our analysis. Note that it contains divergence of the vector $\tau^\alpha$ where components of $\tau^\alpha$ are dynamical and hence they should obey their own equations of motion. In more details, the variation of the action (\ref{Sfinal}) with respect to $\zeta,\beta,\lambda$ and $\tau^\alpha$ imply following set of equations of motion 
\begin{eqnarray}
&&\beta-R=0 \ , \quad \zeta \sqrt{-g}+c\partial_\alpha\tau^\alpha=0 \ , \nonumber \\
&&\sqrt{-g}-\partial_\alpha\tau^\alpha=0 \ , \quad 
-\partial_\alpha\beta+\partial_\alpha[\frac{\lambda}{(\partial_\beta\tau^\beta)^2}(\sqrt{-g}-\partial_\beta\tau^\beta)^2]+\nonumber \\
&&+2\partial_\alpha[\frac{\lambda}{\partial_\beta\tau^\beta}
(\sqrt{-g}-\partial_\beta\tau^\beta)]=0 \ . 
\nonumber \\
\end{eqnarray}
Note that thanks to the previous equations we obtain an important result
\begin{equation}
	\partial_\alpha R=0 \ . 
\end{equation}
In other words scalar curvature has to be constant on-shell.
Let us now presume that Rastall gravity couples to matter which 
is described by the matter action $S_M$. Then the action 
for whole system is $S=S_R+S_M$ and consequently
the variation of the action with respect to $\delta g^{\mu\nu}$ gives
\begin{eqnarray}
	(1-\zeta)G_{\mu\nu}+ (\nabla_\mu\nabla_\nu \zeta-g_{\mu\nu}\Box \zeta )
	-\frac{1}{2}
	g_{\mu\nu}\zeta \beta-\frac{1}{2}\lambda g_{\mu\nu}
	(\sqrt{-g}-\partial_\alpha\tau^\alpha)=T_{\mu\nu} \ , \nonumber \\
\end{eqnarray}
 where we defined stress energy tensor $T_{\mu\nu}$ as
\begin{equation}
	T_{\mu\nu}=-\frac{1}{\sqrt{-g}}\frac{\delta S_M}{\delta g^{\mu\nu}} \ . 
\end{equation}
Now inserting $\zeta=-c, \beta=R$ to the equation above we obtain equations of motion of Rastall gravity
\begin{equation}\label{Rastallcon}
(1+c)(R_{\mu\nu}-\frac{1}{2}g_{\mu\nu}R)+\frac{1}{2}cg_{\mu\nu}R
=T_{\mu\nu}
\end{equation}
Now if we use identity $\nabla^\mu G_{\mu\nu}=0$ that in case
of the condition $R=\mathrm{const}$ implies 
\begin{equation}
	\nabla^\mu R_{\mu\nu}=0 \ 
	\end{equation}
we obtain from (\ref{Rastallcon}) that the stress energy tensor is conserved 
\begin{equation}
	\nabla^\mu T_{\mu\nu}=0 \ . 
\end{equation}
On the other hand taking the trace of the equation (\ref{Rastallcon}) we obtain 
\begin{equation}\label{TR}
	T=(c-1)R=\mathrm{const} \ . 
\end{equation}
In order to be more explicit let  us presume that the matter has the form of perfect fluid with stress energy tensor
in the form
\begin{equation}\label{SET}
	T_{\mu\nu}=(\rho+p)u_\mu u_\nu+p g_{\mu\nu}  \ , 
\end{equation}
where $\rho$ and $p$ are energy density and pressure of perfect fluid respectively which are related by equation of state $p=p(\rho)$. Further, $u_\mu=g_{\mu\nu}u^\nu$ is 
four-velocity of fluid that obeys on-shell condition $u_\mu u_\nu g^{\mu\nu}=-1$. Then taking the trace of the stress energy tensor (\ref{SET}) we obtain 
\begin{equation}
T=-\rho+3p \ .  
\end{equation}
According to the equation (\ref{TR}) the trace of the stress energy tensor
should be constant which is not possible to satisfy for general fluid except in  the case
of radiation when $T=0$. In this case (\ref{TR}) implies that $R=0$.
On the other hand  solutions of Rastall gravity with $R=0$ are equivalent to solutions of General Relativity that satisfy  the
same condition.

In this short note we wrote covariant form of Rastall gravity action as  was proposed 
\cite{Golovnev:2023evp}. We showed that requirement of covariance
implies that action depends on new four vector which is dynamical so that corresponding equations of motion impose strong condition on the scalar curvature. We mean that this 
result again shows that it is difficult to construct
covariant form of action for Rastall gravity.
\\
\\
{\bf Data Availability Statement}: No Data associated in the manuscript
\\
\\
{\bf Declaration and  Acknowledgment:}
The author declares that he has no conflict of interest.
This work  was
supported by the Grant Agency of the Czech Republic under the grant
GA26-22343S.


\end{document}